\newcommand{\pa}{\partial}
\newcommand{\zb}{{\bar z}}
\newcommand{\Lc}{{\cal L}}
\newcommand{\omegav}{{\vec{\omega}}}
\newcommand{\uv}{{\vec{u}}}
\newcommand{\nablav}{{\vec{\nabla}}}
\newcommand{\Xv}{{\vec{X}}}
\newcommand{\xv}{{\vec{x}}}
\newcommand{\Eq}[1]{Eq.(\ref{#1})}
\newcommand{\PB}[2]{\left[ #1, #2 \right]_{PB}}
\newcommand{\DB}[2]{\left[ #1, #2 \right]_{D}}
\newcommand{\der}[2]{\frac{\partial{#1}}{\partial #2}}
\newcommand{\jfm}[3]{{\sl J. Fluid. Mech.} {\bf #1}
(19#2) #3}
\newcommand{\pl}[3]{{\sl Phys. Lett.} {\bf #1}
(19#2) #3}
\newcommand{\np}[3]{{\sl Nucl. Phys.} {\bf #1}
(19#2) #3}
\newcommand{\cmp}[3]{{\sl Commun. Math. Phys.} {\bf #1}
(19#2) #3}
\newcommand{\mpl}[3]{{\sl Mod. Phys. Lett.} {\bf #1}
(19#2) #3}
\newcommand{\cqg}[3]{{\sl Class. Quautum Grav.} {\bf #1}
(19#2) #3}

\documentstyle[12pt]{article}

\begin{document}
%%%%%%%%%%%%%%%%%%

\begin{titlepage}
\begin{flushright}
{\bf UT-644}\\
{\bf May, 1993}
\end{flushright}
\vskip 15mm
\begin{center}

{\bf Hopf Term,  Loop Algebras and}\\
{\bf Three Dimensional Navier-Stokes Equation }

\vspace{1.5cm}

{\bf Yutaka Matsuo}
\vspace{1.0cm}

{\small  Department of Physics, University of Tokyo\\
Hongo 7-3-1, Bunkyo-ku, Tokyo 113, Japan\\
E-mail: matsuo@danjuro.phys.s.u-tokyo.ac.jp}

\vspace{1.9cm}
\end{center}

\abstract{
The dynamics of the 3 dimensional perfect fluid
is equivalent to the motion of vortex filaments
or ``strings".
We study the
action principle and find that
it is described by the Hopf term
of the nonlinear sigma model.
The Poisson bracket structure is described by
the loop algebra, for example, the Virasoro algebra
or the analogue of O(3) current algebra.
As a string theory, it is quite different from the
standard Nambu-Goto string in its
coupling to the extrinsic geometry.
We also analyze briefly the two dimsensional case
and give some emphasis on the $w_{1+\infty}$ structure.
}
\vfill
\end{titlepage}

\section{Introduction}

As Polyakov emphasized in his recent lecture\cite{P1}, the
string theory\cite{GSW} may be one of the unifying concepts in
physics, which is quite essential to the
understanding  of
QCD, quantum gravity, cosmology or
even three dimensional Ising model.
The application of string theory to
these fields are enormously successful,
due to its robust mathematical
strucure, characterized by
the Virasoro algebra, the Kac-Moody algebras,
BRST symmetry or topological methods.
In some occations, we could derive
 exact results just as the consequence of their
infinite dimensional symmetries.

Some months ago, Polyakov\cite{P2} (See also \cite{CT})
made a bold proposal
that the conformal symmetry may determine the
critical exponent of the two dimensional turbulence
exactly.  In his argument, only some simple hypothesis
together with the conformal invariance gives
an discrete series of permitted spectrum of the exponent of
the turbulence. For three dimensional case,
Migdal\cite{M} proposed to use the loop expansion
method of QCD may give
a nice tool to analyze turbulence.
It seems that serious attempt to use string technique
to the fluid dynamics has just begun.

In this letter, we would like to propose
another ``stringy" viewpoints to the Navier-Stokes equation.
In everyday life, virtually everybody
knowns the existence of
``string"---
the vortex filament. It may be observed everywhere,
in the motion of tornado, cigaret smoke, or in
the coffee cup.
Usually, this extended object is
dissipated after a  while.  However,
in the limit where the viscosity vanishes,
these filaments can keep their identities due to
the conservation of circulation.
In this case, the field theoretical description
of fluid reduces  to the dynamics of vortex filament.
Although the classical motion of this ``string"
has been studied in every detail \cite{V},
we do not seem to have enough
knowledge from the viewpoint of ``string theory".
In this letter,
we derive the action principle and
the Poisson bracket structure for the vortex filament.
The kinetic action does not have the standard form
but is described by the Hopf term of the
$O(3)$ non-linear sigma model\cite{P3}.
Since our action is first order with respect to
time, we need to use the canonical formalism for
the singular system. The algebraic structure,
thus obtained, is given by
loop algebras, such as the Virasoro algebra
or an analogure of the $O(3)$ current algebra.
We leave the application of our formalism in the future issues.

\section{Two-dimensional vortices and $w_{1+\infty}$ algebra}
As a warm-up, let us think of the dynamics of two-dimensional
vortices.  It is interesting in its own sake since
it gives a nice representation of $w_{1+\infty}$ algebra.\cite{APD}
In two dimensions, the Navier-Stokes equation takes the following
form,
\begin{equation}
	\der{\omega}{t}=-\nabla_\alpha G_{\alpha},
	\label{1}
\end{equation}
where $\omega(x)=\epsilon_{\alpha\beta}
\pa_\alpha u_\beta(x)$ is the two dimensional
vorticity field and
\begin{equation}
	G_\alpha(x)=\omega u_\alpha(x)-\nu\pa_\alpha \omega.
	\label{2}
\end{equation}
$\nu$ is the viscocity.

It is very important that
if we start from the delta-function distribution
for the vorticity field,
\begin{equation}
	\omega(x)=\sum_{k=1}^N \Gamma_k \delta(x-
	x_k(t)),
	\label{3}
\end{equation}
it keeps its form under time evolution as long as
the viscocity is vanishing.
The dynamics of the Navier-Stokes equation
(field theory equation) thus
reduces to the motion of vortex ``particles", i.e.
the equation of motion can be written by $x_k(t)$ alone.
To see this, it is convenient to
combine $x$ and $y$ coordinate
by $z=x+iy$ and $\zb=x-iy$. With these variables,
the equation of motion for the vortex center becomes,
\begin{equation}
	\der{z_k(t)}{t}=-\frac{1}{2\pi i}\sum_{j(\neq k)}
	\frac{\Gamma_j}{\zb_k-\zb_j}
	\qquad
	\der{\zb_k(t)}{t}=\frac{1}{2\pi i}\sum_{j(\neq k)}
	\frac{\Gamma_j}{z_k-z_j}.
	\label{4}
\end{equation}
Although these equations seem quite innocent, it is
known that they are far from integrable.
Even for the N=3 case, the motion of the
vortices become chaotic.

These equations reminds us strongly the motion of anyon.
Indeed, the Lagrangian of the system is
in some sense related to
the $U(1)$ Chern-Simon term.  To derive it, let us
discuss its Hamiltonian dynamics. The energy functional
of the system is given by
\begin{eqnarray}
	H & = & \frac{1}{2}\int d^2x \sum_{\alpha=1,2}
	u_\alpha(x)u_\alpha(x)
	\nonumber \\
	 & = & -\frac{1}{8\pi}\sum_{j\neq k}	 \Gamma_j\Gamma_k\ln
	 \left[(z_j-z_k)(\zb_j-\zb_k)\right].
	\label{5}
\end{eqnarray}
The sympletic structure which gives the Hamiltonian
equation of motion is
\begin{equation}
	\PB{f}{g}
	= -2i\sum_{j=1}^N \frac{1}{\Gamma_j}
	\der{(f,g)}{(z_j,\zb_j)}.
	\label{6}
\end{equation}
The equations of motion \Eq{4} are rewritten as
$\der{z_j}{t}=\PB{z_j}{H}$ and
$\der{\zb_j}{t}=\PB{\zb_j}{H}$.
In this system, the coordinates $z$ and $\zb$ are the
canonical conjugate with each other.
The Lagrangian which gives this sympletic structure
and the equation of motion is given by,
\begin{equation}
	\Lc=-\frac{i}{4}\sum_{i=1}^N \Gamma_i
	\der{z_i}{t}\zb_i +\frac{1}{8\pi}
	\sum_{i\neq j}\Gamma_i\Gamma_j\ln
	\left[ (z_i-z_j)(\zb_i-\zb_j)\right].
	\label{7}
\end{equation}

The motion of incompressive fluid gives the diffeomorphism
of two dimensional surface which preserve the area element
(area preserving diffeomorphism=APD).
In the string theory, this symmetry is
sometimes called $w_{1+\infty}$
algebra and plays essential
role in several places.\cite{APD}
Dynamical variables of vortices
give a representation of this algebra
through Poisson bracket. In general, an element of
APD is given by the diffeomorphism,
\begin{equation}
	\delta z=\der{\epsilon(z,\zb)}{\zb}
	\qquad
	\delta \zb=-\der{\epsilon(z,\zb)}{z}.
	\label{8}
\end{equation}
Since
\begin{equation}
	\PB{z_i}{\epsilon(z_j,\zb_)}=\delta_{ij}
	\frac{2}{i\Gamma}\der{\epsilon(z,\zb)}{\zb}
	\qquad
	\PB{\zb_i}{\epsilon(z_i,\zb_i)}=-\delta_{ij}
	\frac{2}{i\Gamma}\der{\epsilon(z,\zb)}{z},
	\label{9}
\end{equation}
the generator of the transformation \Eq{8}
is given by
\begin{equation}
	\hat{\epsilon} = \frac{i}{2}
	\sum_{j=1}^N \Gamma_j \epsilon(z_i,\zb_i)
	  =  \frac{i}{2}\int d^2 x \omega(x)
	 \epsilon(x).
	\label{10}
\end{equation}
In a sense, it gives the relation
between the operator of the quantum mechanics $z_i$
and $\zb$ and the operator of the quantum field theory,
$\omega$.  This correspondence reminds us of the
the collective field theory method.\cite{JS}\footnote{
I would like to thank Dr. S. Iso
for pointing out this fact.}
The Poisson bracket between
$\omega$ is
\begin{equation}
	\PB{\omega(x)}{\omega(y)}=
	-2i\sum_{i,j}\epsilon_{ij}\der{}{x^i}
	\der{}{y^j}(\omega(y)\delta(x-y)).
	\label{11}
\end{equation}
This is a standard Poisson bracket which defines
the APD.

Recently, Polyakov\cite{P2} proposed an approach to
the two dimensional turbulence based on the
conformal field theory.  In his method, $\psi=
\Delta^{-1}\omega$ is identified with a primary
field in a minimal model of CFT.
The operator production of the $\psi$ field
is essential to deal with the non-linear term
in Navie-Stokes equation.  The proposed OPE for
$\psi$ field is that
\begin{equation}
	[\psi] \times [\psi] \sim [\phi] + \cdots.
	\label{12}
\end{equation}
On the right hand side of this OPE, there should not
appear the $\psi$ field itself.
Since $\psi$ field is the pseudoscalar, it should
be true as long as the symmetry of the system is
$Vir$ $\times$ $\overline{Vir}$.  This statement is
somehow in contradiction to our Poisson bracket
\Eq{11}. This is possible since our symmetry is APD
and the structure constants
themselves become parity-odd.

Of course, it is dangerous to think that
one may directly apply
the $w_{1+\infty}$ symmetry to the two dimensional turbulence.
In that case,  we take the limit when the
Reynolds number approaches infinity.
As a consequence, the viscocity disappears in the final expression.
However, even if the viscocity is infinitely small, there
appears finite amount of outcome.
It is because
 the
essential nature of the turbulence comes from
the balance between the dissipation due to the
viscosity and the enstrophy flow from outside.
As Polyakov\cite{P2} remarked in his paper, the situation
seems to be similar to the anomaly
in the quantum field theory.

In this way, the $w_{1+\infty}$ symmetry may be
broken in the turbulence even if we
take the limit of vanishing visicocity.
It is extremely interesting to investigate how
$w_{1+\infty}$ is broken and possibly the Virasoro structure
emerges.

\section{Action for vortex filament}
For the three dimensional case, the
computation becomes more involved but basical
strategy remains the same.  In this case, instead of
thinking about point vortices, we need to
consider the vortex string.
The Navier-Stokes equation which is convenient
for us is
\begin{equation}
	\der{\omegav}{t}=-\nablav\times\vec{G},\qquad
	\vec{G}=\omegav\times\uv.
	\label{3.1}
\end{equation}
The vortex filament is described by the
vortex field,
\begin{eqnarray}
	\omegav(x) & = & \sum_{i=1}^N
	\Gamma_i\int d\sigma_i
	\der{\Xv_i(\sigma_i,t)}{\sigma_i}
	\delta^{(3)}\left(\xv-\Xv_i(\sigma_i,t)\right)
	\label{3.2} \\
	\omegav & = & \nablav\times\uv
	\nonumber \\
	\uv(x) & = & \sum_{i=1}^N \frac{\Gamma_i}{4\pi}
	\int d\sigma_i \der{\Xv_i}{\sigma_i}
	\times \frac{\xv-\Xv_i}{|\xv-\Xv_i|^3}.
	\label{3.3}
\end{eqnarray}
One puts these formulae into the Navier-Stokes equation.
After some computations, one may find
the time evolution of the ``string field"
$\Xv(\sigma,t)$,
\begin{equation}
	\der{\Xv_j}{\sigma_j}\times
	\der{\Xv_j}{t}=
	\der{\Xv_j}{\sigma_j}\times
	\left(\sum_{k}^N\frac{\Gamma_k}{4\pi}
	\int d\sigma_k \der{\Xv_k}{\sigma_k}
	\times \frac{\Xv_j-\Xv_k}{|\Xv_j-\Xv_k|^3}
	\right).
	\label{3.4}
\end{equation}
This is equivalent to
\begin{eqnarray}
	\der{\Xv_j}{t}&=&\sum_{k}^N\frac{\Gamma_k}{4\pi}
	\int d\sigma_k \der{\Xv_k}{\sigma_k}
	\times \frac{\Xv_j-\Xv_k}{|\Xv_j-\Xv_k|^3}
	+ \alpha \der{\Xv_j}{\sigma_j}\nonumber\\
 &=& \uv(\Xv_j)+\alpha \der{\Xv_j}{\sigma_j}.
	\label{3.5}
\end{eqnarray}
$\alpha$ can not be determined from the Navier-Stokes
equation alone. This term is present because
of the freedom of the parametrization of the string $\sigma_j$.

We claim that the  action of the vortex filament is given by,
\begin{eqnarray}
	S & = & S_0- H
	\nonumber \\
	 S_0 & = & \sum_{j=1}^N
	 \frac{\Gamma_j}{3}\int dt d\sigma_j
	 \Xv_j\cdot\left(\der{\Xv_j}{\sigma_j}\times
	 \der{\Xv_j}{t}
	 \right)\nonumber\\
	 H& = &
\frac{1}{2}\int d^2x |\uv(x)|^2
\nonumber\\
 & = & \frac{1}{8\pi}
	 \sum_{jk}\Gamma_j\Gamma_k
	 \int d\sigma_j \int d\sigma_k
	 \left(\der{\Xv_j}{\sigma_j}\cdot
	 \der{\Xv_k}{\sigma_k}
	 \right)
	 \frac{1}{|\Xv_j-\Xv_k|}.
	\label{3.6}
\end{eqnarray}
As far as we know, this action seems to be new.
The second part of the action is
identical to the Hamiltonian of the system.
Together with its variation,
\begin{equation}
	\delta H = \frac{1}{4\pi}
	\sum_{jk}\Gamma_j\Gamma_k
	\int d\sigma_j\int d\sigma_k
	\delta \Xv_j\cdot\left(\der{\Xv_j}{\sigma_j}\times
	\left( \der{\Xv_k}{\sigma_k}
	\times \frac{\Xv_j-\Xv_k}{|\Xv_j-\Xv_k|^3}
	\right)\right).
	\label{3.7}
\end{equation}
and the variation of the first term,
\Eq{3.6} gives equation of motion in the form
\Eq{3.4}.

The first term of the action ($S_0$) is known
as the Hopf term in the theory of
O(3)-invariant nonlinear sigma model.\cite{P3}
It counts the instanton numbers and describes
the structure of the $\theta$ vacuum.
It is obvious that our $S_0$ has also topological nature.
Unlike the Nambu-Goto string, it does not couple to
the two dimensional metric on the world sheet.
However, it is invariant under the reparametrization of
world sheet variables because of
its Chern-Simon type structure.

This situation strongly
reminds us of the Chern-Simon approach to
the braid group \cite{W1} where
Witten has shown that the Jones
Polynomial can be
described by the correlation functions
of $su(2)$ Chern-Simon theory.
In our case, the fundamental conserved quantities of the
3d Navier-Stokes Equation are the energy and the helicity.\cite{MT}
The latter has definite
topological meaning since it is expressed by,
\begin{equation}
h=\int d^3 x \omegav(x)\uv(x)
 = 2\sum_{i<j}\beta_{ij}\Gamma_i\Gamma_j,
\end{equation}
where
\begin{eqnarray}
\beta_{ij}& = & \frac{1}{4\pi}
\int d\sigma_i \int d\sigma_j
\der{\Xv_i}{\sigma_i}\cdot
\left(\der{\Xv_j}{\sigma_j}
\times
\frac{\Xv_i-\Xv_j}{|\Xv_i-\Xv_j|^3}
\right).
\end{eqnarray}
We remark that $\beta_{ij}$ is the linking number of
two filaments $\Xv_i$ and $\Xv_j$.
Also, the interesting analogy is the Chern-Simon
gravity theory in three dimensions.\cite{W2}
Whereas the three dimensional gravity theory
is the gauge theory of  diffeomorphism of three dimensional
space-time, the Poisson bracket of Navier-Stoke equation
is volume preserving diffeomorphism
\cite{A}

Finally, we would like to make a comment on
the difference between the vortex motion and
that of Nambu-Goto string.
The issue is their coupling to the extrinsic curvature\footnote{
This part is a  short review. Readers
who are familiar with vortex motion can skip the rest of this section}.
To see this, we should first remark that
our formulae Eqs.(\ref{3.4}-\ref{3.7}) are not well-defined
due to the divergence  in the $\sigma$ integral.
For example, the integral which appear in the left
hand side of \Eq{3.5} diverges logarithmically.
We put $j=k$ and write $\sigma_j=\sigma$ and
$\sigma_k=\sigma+\theta$. Collecting the most divergent
peace gives,
\begin{equation}
\left.
\der{\Xv}{t}\right|_{\sigma=\sigma_{0}}
	\approx  \left[\frac{\Gamma}{4\pi}\int
	d\theta \frac{1}{2|\theta|}\right]\frac{
	\der{\Xv}{\sigma_0}\times
	\der{^2\Xv}{\sigma_0}}{|\der{\Xv}{\sigma_0}|^2}.
	\label{3.8}
\end{equation}
In this form, one
 recognizes that one may absorb
the divergent factor by the redernition of time variable,
\begin{equation}
t\longrightarrow \tilde{t}=\Gamma\Lambda t,
\qquad
\Lambda = \frac{1}{4} \log(L/\epsilon),
\end{equation}
where $L$ is the length of the vortex and $\epsilon$ is the
ultraviolet cutoff which may be identified with the diameter of
vortex filament. In the following, we rewrite $\tilde{t}$ as just
$t$ for the simplicity.

The remaining term is directly related to the
extrinsic geometry of the curve.
Since the parametrization is arbitrary, we take
a special choice of $\sigma$,
$
	\left|
	\der{\Xv}{s}
	\right|^2=1,
$
i.e $s$ gives the length of the filament.
With this parametrization, we may use the Frenet-Serret
formula,
\footnote{The Frenet-Serret formula of the world
sheet appeared previously
in the context of W-gravity theory.\cite{GM}
The link of vortex-string with W-gravity seems to be a fascinating
subject.}
\begin{equation}
	\der{\Xv}{s}  =  \vec{t},\quad
	\der{\vec{t}}{s}  =  \kappa(s) \vec{n},
	\quad
	\der{\vec{n}}{s}  =  \tau(s)\vec{b}-\kappa(s)\vec{t},
	\quad
	\der{\vec{b}}{s}  =  -\tau(x)\vec{n}.
	\label{3.10}
\end{equation}
One may readily prove that
\begin{equation}
	\frac{
	\der{\Xv}{s}\times
	\der{^2\Xv}{s}}{|\der{\Xv}{s}|^2}=\kappa\vec{b}
	\qquad\left(=\der{\Xv}{t}\right).
	\label{3.11}
\end{equation}
This is well-known equation which describes that
the vortex filament moves to the binormal direction
with the velocity proportional to its curvature at that point.
Very interestingly, Hashimoto derived
that this equation
is exactly solvable.\cite{V}
It was proved that by changing the
variable,
\begin{equation}
	\psi(s)=\kappa(s)\exp\left(
	i\int_{0}^s \tau(s')ds'
	\right),
	\label{3.12}
\end{equation}
\Eq{3.11} becomes the
non-linear Schr\"odinger equation,
\begin{equation}
	\frac{1}{i}\der{\psi}{t}=
	\der{^2\psi}{s^2}+\frac{1}{2}
	(|\psi|^2+A)\psi,
	\label{3.13}
\end{equation}
which is integrable.

\section{Canonical formalism}
The action derived in the previous section is the first order
in the time variable.  Furthermore, there is
the first class constraint because of the reparametrization
invariance of $\sigma$.  This is an ideal
example to learn the canonical formalism of the
singular system.
In the following, we consider the
special case $N=1$ for the simplicity. The extension to
general case is straightforward.

The canonical conjugate of the
string field is given by,
\begin{equation}
	\Pi_\rho(\sigma, t)\equiv
	\frac{\delta S_0}{\delta \dot{X^\rho}}
	=\frac{\Gamma}{3}\epsilon_{\mu \nu \rho}
	X^\mu \der{X^\nu}{\sigma}.
	\label{4.1}
\end{equation}
On the right hand side, there is no time derivative.
Hence the definition of momentum  already
gives constraints.
Let us denote them as $\phi_\rho$,
\begin{equation}
	\phi_\rho(\sigma,t)
	\equiv
	\Pi_\rho(\sigma)-\frac{\Gamma}{3}\epsilon_{\mu \nu \rho}
	X^\mu \der{X^\nu}{\sigma}\approx 0.
	\label{4.2}
\end{equation}
To see the structure of these constrants, we take the
Poisson bracket of them,
\begin{equation}
	\PB{\phi_\mu(\sigma)}{\phi_\nu(\sigma')}
	= -\Gamma\epsilon_{\mu\nu\rho}
	\der{X^\rho}{\sigma}\delta(\sigma-\sigma')
	\equiv M_{\mu\nu}(X)\delta(\sigma-\sigma').
	\label{4.3}
\end{equation}
Following the general procedure,
we add terms proportional to
the constraint to the Hamiltonian,
\begin{equation}
	\delta H = \sum_{\rho=1}^3f_\rho(X,\Pi)\phi_\rho.
	\label{4.4}
\end{equation}
We need to check whether
the time evolution of $\phi$ with respect to
new Hamiltonian gives rize to
any new constraints.
The Poisson bracket of the
constraint with the extended Hamiltonian is,
\begin{equation}
	\PB{\phi_\rho}{H+\delta H} \approx
	\epsilon_{\rho\mu\nu}\der{X^\mu}{\sigma}
	u^\nu(X(\sigma))
	 -\Gamma \epsilon_{\rho\mu\nu}f^\mu\der{X^\nu}{\sigma}.
	\label{4.5}
\end{equation}
If we choose $f^\mu = \frac{1}{\Gamma}u^\mu(X)$, the right hand
side vanishes weakly.  Hence, there are no
secondary constraints.

Since the rank of matrix $M_{\mu\nu}$ is two,
one linear combination of the constraint
fields $\phi_\mu$  should be the
first class. It corresponds to the
reparametrization of the arbitrary parameter $\sigma$.
Explicitly, we define,
\begin{equation}
	T(\sigma)\equiv -\der{X^\mu}{\sigma}\phi_\mu
	=-\der{X^\mu}{\sigma}\Pi_\mu.
	\label{4.6}
\end{equation}
It satisfies the classical version of the Virasoro
algebra,
\begin{equation}
	\PB{T(\sigma)}{T(\sigma')}
	=2T(\sigma')\der{}{\sigma'}\delta(\sigma-\sigma')
	+\der{T(\sigma')}{\sigma'}\delta(\sigma -\sigma').
	\label{4.7}
\end{equation}
Poisson brackets with other fields are
given by
\begin{eqnarray}
	\PB{T(\sigma)}{X^\mu(\sigma')} & = & \der{X^\mu}{\sigma}
	\delta(\sigma -\sigma').
	\label{4.8} \\
	\PB{T(\sigma)}{\Pi_\mu(\sigma')} & = &
	-\Pi_\mu(\sigma)\der{}{\sigma}\delta(\sigma-\sigma').
	\label{4.9} \\
	\PB{T(\sigma)}{\phi_\mu}& = & -\phi_\mu(\sigma)
	\der{}{\sigma}\delta(\sigma - \sigma').
	\label{4.10}
\end{eqnarray}
The last equation shows that $T(\sigma)$ is indeed
the first class constraint.

In order to define the Dirac bracket, we need to
make gauge fixing of the parametrization
$\sigma$.  A natural choice is to use the length of
the vortex filament as $\sigma$,
\begin{equation}
	\chi\equiv\left|\der{X}{\sigma}
	\right|^{2}-1\approx 0.
	\label{4.11}
\end{equation}
The Poisson brackets of $\chi$ with other constraints
are
\begin{equation}
	\PB{\phi_\mu(\sigma)}{\chi(\sigma')}
	=\der{X^\mu}{\sigma'}\partial_{\sigma'}
	\delta(\sigma-\sigma')
	\label{4.12}
\end{equation}
Constraints $\phi$ together with the gauge fixing
condition $\chi$ becomes second class.
In general, if the second class constraint
$\xi_\alpha$ satisfies Poisson bracket,
$\PB{\xi_\alpha}{\xi_\beta}=C_{\alpha\beta}$,
the Dirac bracket is given by
\begin{equation}
	\DB{F}{G}=\PB{F}{G}-\sum_{\alpha,\beta}
	\PB{F}{\xi_\alpha}\left( C^{-1}\right)^{\alpha
	\beta}\PB{\xi_\beta}{G}.
	\label{4.13}
\end{equation}
In our case, from \Eq{4.3} and \Eq{4.12}
the matrix $C$ is given by
\begin{equation}
	C_{AB}(\sigma-\sigma')
	=\left(
	\begin{array}{cccc}
		0 & a_3 & -a_2 & b_1  \\
		-a_3 & 0 & a_1 & b_2  \\
		a_2 & -a_1 & 0 & b_3  \\
		-b_1 & -b_2 & -b_3 & 0
	\end{array}
	\right)
	\qquad
	\left\{
	\begin{array}{l}
		a_i=-\Gamma\der{X^i}{\sigma}\delta(\sigma-\sigma')  \\
		b_i=-\der{X^i}{\sigma'}\der{}{\sigma'}
		\delta(\sigma-\sigma')
	\end{array}
	\right.\; .
	\label{4.14}
\end{equation}
The index $A$ stands for $\phi^A$ for $A$=1,2,3, and $\chi$
for $A=4$.
Because of the constraint $\chi\approx 0$, the inverse
of this matrix becomes very simple,
\begin{equation}
	(C^{-1})^{AB}(\sigma-\sigma')
	=\left(
	\begin{array}{cccc}
		0 & c_3 & -c_2 & d_1  \\
		-c_3 & 0 & c_1 & d_2  \\
		c_2 & -c_1 & 0 & d_3  \\
		-d_1 & -d_2 & -d_3 & 0
	\end{array}
	\right)
	\qquad
	\left\{
	\begin{array}{l}
		c_i=-\frac{1}{\Gamma}
		\der{X^i}{\sigma}\delta(\sigma-\sigma')  \\
		d_i=\der{X^i}{\sigma}
		\theta(\sigma-\sigma')
	\end{array}
	\right.\; ,
	\label{4.15}
\end{equation}
with $\theta$ is the step function.
The Dirac bracket may be computed by using the general
formula \Eq{4.13} and, for special case, it gives,
\begin{equation}
	\DB{X^\mu(\sigma)}{X^\nu(\sigma)}
	= \frac{1}{\Gamma}\epsilon_{\mu\nu\rho}
	\der{X^\rho}{\sigma}\delta(\sigma
	-\sigma').
	\label{4.16}
\end{equation}
This algebra resembles $O(3)$ current algebra.
However, there appear the derivative of $\Xv$
instead of $\Xv$ itself. Furthermore, we should not
forget that there is a constraint \Eq{4.11}.

\section{Discussion}

There are several issues that we could not discuss well
in this letter. Let us make a list to illustrate
the future issues.
\begin{itemize}
\item {\bf Relativistic Formulation:\hskip 3mm }
In this letter, our consideration is restricted to
the non-relativistic situation.  Because of this restriction,
only one Virasoro algebra could appear.  If we formulate
it relativistically, we expect we have
the Virasoro algebras both
in right and left moving sectors.
\item {\bf Quantization and BRST
formalism:\hskip 3mm}  Since our algebra
\Eq{4.11} is slightly different from the current algebra,
the quantization and the BRST analysis seems a little bit non-trivial.
Anyway, what is ``the critical dimension" of our string?

\item {\bf Description of Turbulence:\hskip 3mm}
As Migdal observed\cite{M},
there seems to be a relation between the viscocity and the
plank constant. In this sense, the quantized version of our
analysis should give some aspects of the turbulence.

\item {\bf Relation with the
quantum gravity:\hskip 3mm}
As we discussed in section 3,
there are some analogies between
three dimensional gravity theory
and the Navier-Stokes equation.
It seems very important to investigate
their relation more explicitly.
Also, in two dimensions, the area preserving
diffeomorphism describe some
aspect of the quantum gravity\cite{APD}.
The appearance of APD in two dimensional Navier Stokes
seems to establish another link.

\end{itemize}

\vskip 10mm
{\em The author would like to
thank H. C\^ateau, T.Eguchi, K. Fujikawa,
H. Kawai K. Ogawa and Q.-H. Park for their interest
on this work.
Especially, we are very obliged to
S. Iso for critical discussions
on two-dimensional vortices and
$w_{1+\infty}$ algebra and to
M.Umeki for comments and showing relevant references.}

\vskip 10mm
{\bf Note Added}: \hskip 5mm After we submitted this paper,
we learned that substantial part of this work have been done
previously in beautiful papers \cite{RR}, \cite{L}.
See also \cite{A} where some material on two dimensional
case was discussed. We would like to
give our thanks to Drs. T. J. Allen, M. Genna, S. Yahikozawa
for indicating these references.

\end{document}